\documentclass[runningheads]{llncs}
\usepackage{makecell}
\usepackage{threeparttable}
\usepackage{amsmath}
\usepackage{wasysym}
\usepackage{url}
\usepackage{graphicx}
\usepackage{algorithm}
\usepackage{hyperref}
\usepackage{algorithmic}
\usepackage{booktabs}

\usepackage{lipsum}

\usepackage[underline=false]{pgf-umlsd}
\usetikzlibrary{calc}
\usetikzlibrary{arrows.meta}

\begin{document}
\title{Remote Credential Management with Mutual Attestation for Trusted Execution Environments}
\titlerunning{Remote Credential Management with Mutual Attestation for TEEs}
%

\author{Carlton Shepherd, Raja N. Akram, \and Konstantinos Markantonakis}
\authorrunning{Shepherd et al.}
%
\institute{Information Security Group, Royal Holloway, University of London, Egham, UK\\
\email{\{carlton.shepherd.2014, r.n.akram, k.markantonakis\}@rhul.ac.uk}}
\maketitle              
\begin{abstract}

Trusted Execution Environments (TEEs) are rapidly emerging as a root-of-trust for protecting sensitive applications and data using hardware-backed isolated worlds of execution.   TEEs provide robust assurances regarding critical algorithm execution, tamper-resistant credential storage, and platform integrity using remote attestation.   However, the challenge of remotely managing credentials \emph{between} TEEs remains largely unaddressed in existing literature.  In this work, we present novel protocols using mutual attestation for supporting four aspects of secure remote credential management with TEEs: \emph{backups}, \emph{updates}, \emph{migration}, and \emph{revocation}.  The proposed protocols are agnostic to the underlying TEE implementation and subjected to formal verification using Scyther, which found no attacks. 

\end{abstract}

\section{Introduction}
\label{sec:intro}

Trusted computing offers robust, on-device protection of security-critical data and the ability to securely report evidence of platform integrity, which has culminated in efforts such as the Trusted Platform Module (TPM).  Until recently, however, such technologies were relatively restricted: neither arbitrary application execution nor secure input/output (I/O) are realisable with TPMs without substantially increasing the hardware-software Trusted Computing Base (TCB), say, through the use of virtual machines \cite{greene2012intel}.  
Trusted Execution Environments (TEEs), discussed further in Section \ref{sec:tees}, have emerged as the forerunner for addressing these shortcomings, particularly for constrained devices~\cite{sadeghi2015security}.   Unlike TPMs, TEEs provide hardware-enforced isolated execution of critical applications and data on the same underlying hardware. TEEs aim to thwart sophisticated software adversaries from a conventional Operating System (OS) irrespective of its protection mode, e.g\ Rings 0--3.   Modern Intel and ARM chipsets offer Intel Software Guard eXtensions (SGX) and ARM TrustZone respectively for instantiating a TEE from the CPU or System-on-Chip (SoC).  

Despite widespread availability, managing TEE data credentials throughout their life-cycle has received little attention by the community.  Such credentials, whether derived from a public-key certificate, password or another value, are typically used to authenticate sensitive actions and transmitted data.  Challenges arise, however, when credentials require migrating, revoking, updating or backing-up in a secure and trusted manner with bi-directional assurances between both end-points.  
Firstly, large numbers of TEEs must be administered, thus limiting the feasibility of human intervention, potentially over a multitude of communication mediums.   Secondly,  heterogeneous TEEs must be accommodated: Intel SGX, for example, is confined to Intel CPUs on more powerful devices, while ARM TrustZone is limited to ARM-based SoCs.
Hence, for the first time, we address four key challenges when managing heterogeneous TEE credentials over its lifetime with bi-directional trust assurances for remote \emph{migration} (Section \ref{sec:mig}), \emph{revocation} (Section \ref{sec:revoc}), \emph{backups} (Section \ref{sec:backup}), and \emph{updates} (Section \ref{sec:update}).  
This paper presents the following contributions:
\begin{itemize}
\item An examination of existing smart card and TPM work relating to each credential management challenge and their applicability to TEEs.
\item A suite of proposed protocols for facilitating TEE credential management with mutual attestation.  The protocols are agnostic of the TEE and communication medium, and employ a Trusted Service Manager (TSM) in line with the GlobalPlatform TEE specifications~\cite{gp:tmf}.
\item The proposed protocols are subjected to formal symbolic verification using Scyther, which found no attacks.  We publicly release the verification specifications for further research\footnote{Available online at: \url{https://cs.gl/extra/wistp18-scripts.zip}}.
\end{itemize}

\section{Trusted Execution Environments (TEEs)}
\label{sec:tees}

GlobalPlatform defines a TEE as an isolated execution environment that \emph{``protects from general software attacks, defines rigid safeguards as to the data and functions a program can access, and resists a set of defined threats''}~\cite{gp:tee}.  TEEs aim to isolate applications from integrity and confidentiality attacks from the untrusted OS, e.g. Android, or Rich Execution Environment (REE), by allocating distinct memory regions with accesses controlled by hardware.  
We summarise the foremost commercial TEEs for Intel and ARM chipsets; the reader is referred to \cite{shepherd:secure} for a detailed survey of secure and trusted execution environments. 

\textbf{Intel Software Guard eXtensions (SGX)} is an extension to the x86-64 instruction set that enables the creation of per-application `enclaves'.  Enclaves reside in isolated memory regions within RAM with accesses mediated by the CPU, which is considered trusted~\cite{sgx:costan}. 
Secure storage is provided via the sealing abstraction, where data is encrypted to the untrusted world using a key derived from a processor-specific Storage Root Key (SRK).  Enclave- or author-specific keys can be derived; that is, respectively, binding data to only that enclave, or from an ID string to preserve persistence between enclaves from the same author.  
Remote attestation enables remote verification of enclaves and secret provisioning using the Enhanced Privacy ID (EPID) scheme~\cite{brickell:epid}, which authenticates enclave integrity measurements without revealing the CPU's identity.  

\textbf{GlobalPlatform (GP) TEE with ARM TrustZone} maintains two worlds for all trusted and untrusted applications. A TEE kernel is used for scheduling, memory management, cryptographic methods and other OS functions, while user-mode TEE Trusted Applications (TAs) access OS functions exposed by the GP TEE Internal API. 
The GP TEE Client API \cite{gp:tee} defines the interfaces for communicating with TAs from the REE. 
The predominant method for instantiating the GP TEE is with ARM TrustZone, which enables two isolated worlds to co-exist in hardware using two virtual cores for each world per physical CPU core and an extra CPU bit (NS bit) for distinguishing REE/TEE execution modes.  TrustZone provides secure I/O with peripherals connected over standard interfaces, e.g. SPI and GPIO, by routing interrupts to the TEE kernel using the TrustZone Protection Controller (TZPC) for securing on-chip peripherals from the REE, and the Address Space Controller (TZASC) for memory-mapped devices. Both TZASC and TZPC utilise the NS bit for access control.
The GP TEE implements secure storage using the sealing abstraction described previously, or to TEE-controlled hardware, e.g. Secure Element (SE).

\begin{figure}[t]
\centering
\includegraphics[width=0.95\linewidth]{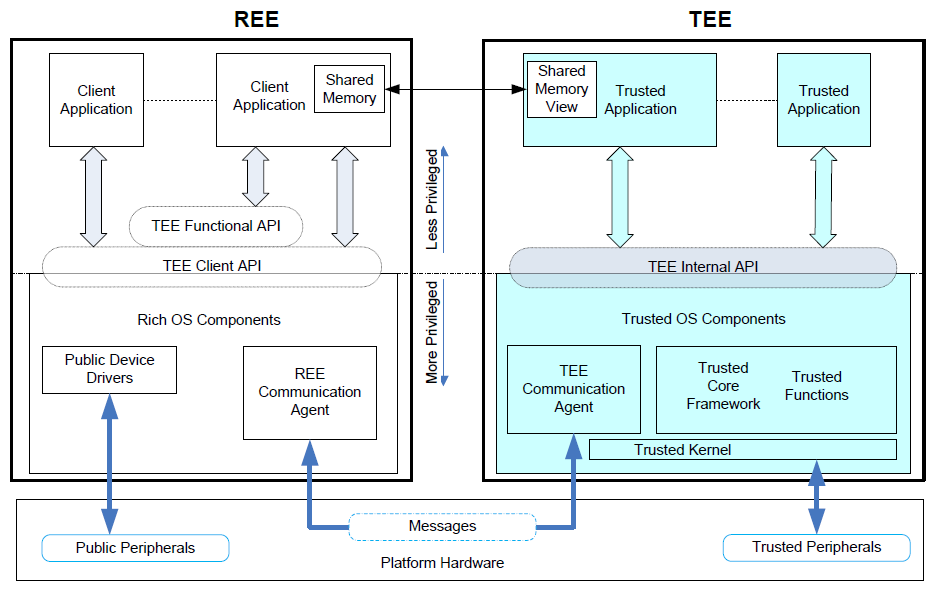}
\caption{GlobalPlatform TEE system architecture \cite{shepherd:secure}.}
\label{fig:gptee}
\vspace{-0.6cm}
\end{figure}

\textbf{Credential Management.} 
 Security credentials are the evidence that a communicating party possesses for accessing privileged data and services.  
Credentials are typically programmed initially into a TEE during the personalisation phase following the procurement of the SoC and TEE software, but prior to deployment.  After this, a Trusted Service Manager (TSM) -- incorporated into the device manufacturer or outsourced -- is responsible for maintaining the TEE, its TAs and on-board credentials thereafter. 
We define TEE credentials as \emph{the set, $C$, of key material, certificates and other authentication data issued by $TSM$ that is provisioned into a TA}.  TEE credentials may also comprise a key derived from a password-based key derivation function using a password from an operator, or encapsulated by biometrics, e.g. iris and fingerprint, or a behavioural model that maps device continuous data to authentication states~\cite{shepherd2017towards}.

\subsection{Credential Management: Security and Functional Requirements}
\label{sec:reqs}

The GP Trusted Management Framework (TMF) does not stipulate particular secure channel protocols, but only that the TSM and TEE should mutually authenticate over a channel that preserves the \emph{``integrity and the confidentiality of the exchanges,''} and addresses replay attacks against a Dolev-Yao adversary~\cite{gp:tmf}.     
These basic requirements omit desirable features identified in existing literature~\cite{akram:trusted,greveler,shepherd:ares}, such as assurances that the target TEE is authentic and integral.  This is typically realised using Remote Attestation (RAtt) where, firstly, system measurements are taken at boot-time or on-demand, which are collected and signed by a trusted measurer under a device-specific key; RAtt protocols subsequently transmit the measurements over a secure channel to a remote verifier, who evaluates the platform's integrity based on these values. 

In sensitive deployments, mutually authenticating \emph{both} end-points is useful during TEE-to-TEE communication; for example, between uploading backups from a GP TEE to a cloud-based backup enclave using Intel SGX.  Here, RAtt protocols can be conducted independently for each end-point, or using a \emph{mutual attestation} protocol wherein both parties are attested in a single protocol instance.  We refer to such mutual attestation protocols, e.g. \cite{shepherd:ares}, as providing a Secure and Trusted Channel Protocol (STCP) in this work.  
We now formalise the baseline security and functional requirements from issues raised in related work and those stipulated in the GlobalPlatform specifications:
\begin{enumerate}
\item[S1)] \emph{Mutual key establishment}: a shared secret key is established for communication between the two entities.
\item[S2)] \emph{Forward secrecy}: the compromise of a particular session key should not affect past or subsequent protocol runs.
\item[S3)] \emph{Trust assurance}: the proposal shall allow third-parties to verify the target platform's integrity prior to credential transmission.
\item[S4)] \emph{Mutual trust verification}: both end-points shall successfully attest the state of the other before permitting the establishment of a secure channel.
\item[S5)] \emph{Mutual entity authentication}: each communicating end-point shall authenticate the other's identity to counter masquerading attempts.
\item[S6)] \emph{Denial of Service (DoS) resilience}: resource allocation shall be minimised at both end-points to prevent DoS conditions from arising.
\item[S7)] \emph{Key freshness}: the shared key shall be fresh to the session in order to prevent replay attacks.
\item[F1)] \emph{Avoidance of additional trust hardware}: the protocol shall avoid the need for additional security hardware, e.g. TPMs and SEs, other than the TEE. 
\item[F2)] \emph{TEE agnosticism}: the protocols shall remain agnostic of the underlying TEE architecture to facilitate interoperability.
\end{enumerate}

\textbf{Setup Assumptions.}
A public-key infrastructure is assumed in which a Certificate Authority (CA) issues certificates to the TSM, TAs, and backup (BA), revocation (RA) and maintenance (MA) authorities used for managing backups, revoking credentials and physically maintaining devices respectively.   The TEEs themselves are assumed to be trusted and to possess certified, device-specific attestation and command keys for signing quotes and requests to the TSM.  Quotes are a widely-used remote attestation abstraction for TPMs and TEEs, comprising the TEE's identity and the platform integrity measurements collected by a TEE-resident trusted measurer.  The resulting quote is signed using the attestation key and transmitted to the remote verifying authority.   The credentials are assumed to be securely stored within the TEE, usually performed by encrypting them under a device-specific SRK, as well as secure means of random number generation and key derivation.     

\section{Migration}
\label{sec:mig}
TEE migration is the process of transferring and re-provisioning credential data from $TA_A$ to $TA_B$ in distinct TEEs.  Migration is crucial in preserving credentials during a device replacement or relocation, where credentials can be remotely transferred without incurring reinitialisation costs.   Migrating credentials across TEEs has already attracted some attention in related literature~\cite{arfaoui,kost}.  We summarise these schemes and their contributions.

Arfaoui et al.~\cite{arfaoui} present a privacy-preserving protocol for migrating credentials between GlobalPlatform TEEs.  A PKI-based protocol is proposed for authorising the credential transfer between the TEE, TSM and the TEE's service providers, each of whom controls a set of TAs in a Security Domain (SD).  After authorisation, a second protocol is used to transfer data between the SDs by each service provider using a PKI- or password-based authenticated key exchange.  Both protocols are subjected to formal verification using the AVISPA analysis tool.  While the authors note the importance of remote attestation during TSM-TEE authorisation, it is not presented or verified as part of the protocol; it is also omitted during the credential transfer process between the TEEs.  Moreover, \emph{mutual} trust assurances between the TSM and the TEEs is not discussed.

Kostiainen et al.~\cite{kost} tackle migration for TEE open credential platforms where service providers can provision arbitrary credentials, say, for virtualised access control cards.  The authors propose encrypting and backing-up credentials on a trusted server using a tokenised password known only to the user.  The credentials are migrated by re-entering the password, which is re-tokenised on the receiver device, and transmitted and verified by the backup server that releases the encrypted credentials.  However, like \cite{arfaoui}, the proposal lacks trust assurances between the TSM and both TEEs, nor is it subjected to formal analysis.



\subsection{Proposed High-level Migration Procedure}

\begin{figure}[t]
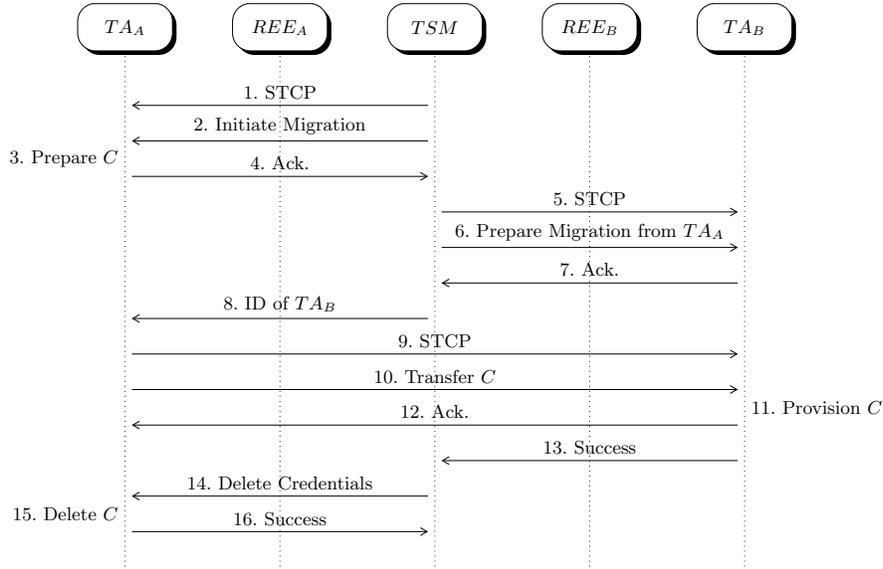

	\centering
	\resizebox{\linewidth}{!}{%
		\begin{sequencediagram}
		    \tikzstyle{inststyle}+=[rounded corners=3mm]
		    \newinst{TA1}{$TA_A$}
            \newinst[1]{REE1}{$REE_A$}
			\newinst[1]{TSM}{$TSM$}
			\newinst[1]{REE2}{$REE_B$}
			\newinst[1]{TA2}{$TA_B$}
            \mess{TSM}{1. STCP}{TA1}
            \mess{TSM}{2. Initiate Migration}{TA1}
            \node[below=3mm,anchor=east](t1) at (mess to) {3. Prepare $C$};
            \mess{TA1}{4. Ack.}{TSM}
            \mess{TSM}{5. STCP}{TA2}
            \mess{TSM}{6. Prepare Migration from $TA_A$}{TA2}
            \mess{TA2}{7. Ack.}{TSM}
            \mess{TSM}{8. ID of $TA_B$}{TA1}
            \mess{TA1}{9. STCP}{TA2}
            \mess{TA1}{10. Transfer $C$}{TA2}
            \node[below=3mm,anchor=west](t1) at (mess to) {11. Provision $C$};
            \mess{TA2}{12. Ack.}{TA1}
            \mess{TA2}{13. Success}{TSM}
            \mess{TSM}{14. Delete Credentials}{TA1}
            \node[below=3mm,anchor=east](t1) at (mess to) {15. Delete $C$};
            \mess{TA1}{16. Success}{TSM}
        \end{sequencediagram}
	}
	\caption{Proposed TEE credential migration procedure.}
	\label{fig:migration}
	\vspace{-0.5cm}
\end{figure}

Credentials must be deleted on the device from which they are migrated, while transferring them over a secure channel with mutual trust assumptions.  In Figure \ref{fig:migration}, we show how migration can be performed between remote TAs accounting for the shortcomings in related work, which comprises the following messages:
\begin{enumerate}
\item A mutual remote attestation protocol~\cite{akram:trusted,greveler,shepherd:ares} is executed between $TSM$ and $TA_{A}$ to bootstrap a secure and mutually trusted channel (STCP).
\item $TSM$ transmits the begin migration command to $TA_{A}$.
\item $TA_{A}$ unseals credentials, $C$, from its secure storage for transmission.
\item $TA_{A}$ acknowledges to $TSM$ that the credentials were unsealed successfully.
\item A separate STCP instance is executed between $(TSM, TA_{B})$.
\item $TSM$ instructs $TA_{B}$ to prepare for credential provisioning.
\item $TA_{B}$ acknowledges to $TSM$ that it is ready to receive credentials.
\item $TSM$ transmits the ID of $TA_{B}$ to $TA_{A}$, e.g. IP address, to which to transmit the unsealed credentials.
\item An STCP is formed using mutual remote attestation between $TA_{A}$ and $TA_{B}$.
\item The credential transfer occurs between $TA_{A}$ and $TA_{B}$.
\item The transferred credential is provisioned into the secure storage of $TA_{B}$.
\item[12--13.] $TA_{B}$ acknowledges its provisioning success to $TA_A$ and $TSM$.
\item[14.] $TSM$ instructs $TA_{A}$ to delete its migrated credential(s).
\item[15--16.] $TA_{A}$ deletes $C$ and acknowledges its success to $TSM$.
\end{enumerate}

The high-level procedure uses three STCPs with mutual attestation between $(TSM,TA_{A})$, $(TSM, TA_{B})$ and $(TA_{A}, TA_{B})$, thus addressing the absence of bi-directional trust assurances in existing work.  The proposal avoids unnecessarily exposing credentials to the TSM by transmitting data directly between the mutually authenticated TAs. Implicitly, the protocol avoids specifying TEE-specific functionalities
; rather, for F2 (TEE agnosticism), we abstract the protocol appropriately to allow migrations between heterogeneous TEEs by allowing either a GP TEE application or Intel SGX enclave to act as either $TA$.  For TEE-specific implementation guidance, the reader is referred to existing work such as the GlobalPlatform TMF specifications~\cite{gp:tmf}, and the work by Arfaoui et al.~\cite{arfaoui} for managing and authorising SDs on the GP TEE.  
In Section \ref{sec:protocols}, we specify the protocols and procedures formally, and detail an enhanced mutual attestation protocol for STCP for satisfying the remaining requirements from Section \ref{sec:reqs}.

\section{Revocation}
\label{sec:revoc}
Credentials are typically revoked when they reach the end of their predefined lifespan as part of a key rotation policy; if the OEM discovers a vulnerability in the TA or TEE kernel code, and the credentials were potentially compromised; or the device is retired from service, say, due to obsolescence.
Revocation has attracted much attention in related TPM and smart card literature.  Chen and Li~\cite{chen} address credential revocation in TPM 2.0 Direct Anonymous Attestation (DAA).  Like conventional group signatures, DAA allows the signer to demonstrate knowledge of its individual private key corresponding to the group's public key; however, this complicates revocation because the signer's identity is not revealed
, even to the group manager.  The authors review two solutions: \emph{rekey-based}, where the issuer regularly updates its public key (which may or may not include its corresponding secret key), allowing only legitimate non-revoked signers to update their credentials accordingly; and \emph{Verifier-Local Revocation (VLR)}, where the verifier inputs a revocation list ($RL$), to the DAA's verification function and accepts only signatures from signers $S \notin RL$.  

Lueks et al.~\cite{lueks} address revoking attribute-based credentials (ABCs) for smart cards anonymously.  
Here, the RA possesses an RL of anonymous revocation values, $g^{r}_{\epsilon,v}$, submitted by the user or verifier (user- and system-instantiated revocation), where $r$ is the revocation value in the user's credential.  A revocation `epoch', $\epsilon$, corresponding to a time period, is used to provide unlinkability by re-computing and re-sending the new valid RLs to the verifiers at each epoch; that is, $RL_{\epsilon,V}=sort(\{g^{r}_{\epsilon,V}\mid r\in MRL\})$, where $MRL$ is the master revocation list.  Using bloom filters, this occupies only 4--8MB for $2^{21}$ revoked credentials depending on the chosen probability tolerance.  

Katzenbeisser et al.~\cite{katz} propose revocation for TPM 1.2 using blacklisting and whitelisting.  
For blacklisting, a list of revoked keys, $BL$, is ordered into a hash chain and encrypted under the TPM's Storage Root Key (SRK); the final hash chain value is stored in a TPM register to maintain integrity.   Before loading a key, $k'$, each $k_i \in BL$ is fed sequentially into the TPM, where it is decrypted, and $k'\stackrel{?}{=} k_i$ is tested.  Whitelisting incrementally creates keyed hashes of each permitted key under the TPM's SRK and internal secure counter representing the whitelist's `version'. 
A key is valid \emph{iff} the keyed hash counter value matches the TPM's internal counter.  Revocation is performed by incrementing the TPM's counter and updating all non-revoked hashes with the new value.


\subsection{Proposed High-level Revocation Procedure}
\begin{figure}
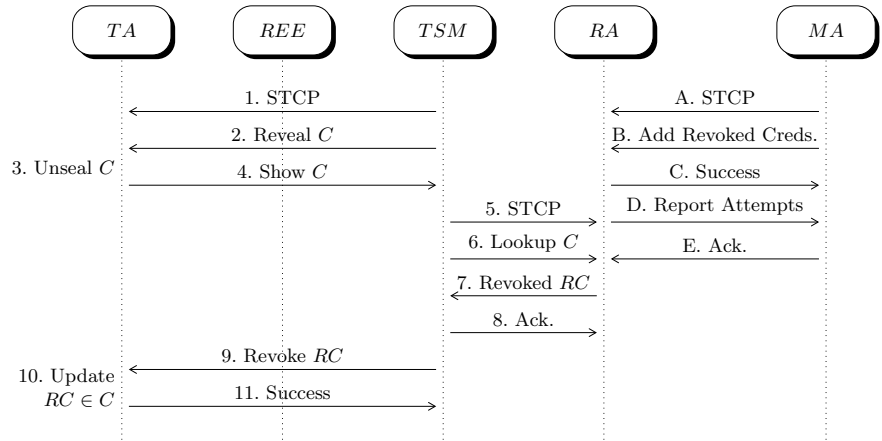

	\centering
	\resizebox{\linewidth}{!}{%
		\begin{sequencediagram}
		    \tikzstyle{inststyle}+=[rounded corners=3mm]
		    \newinst{TA}{$TA$}
            \newinst[1]{REE}{$REE$}
			\newinst[1]{TSM}{$TSM$}
			\newinst[1]{RA}{$RA$}
			\newinst[2]{MA}{$MA$}

			\mess{MA}{A. STCP}{RA}
            \mess{MA}{B. Add Revoked Creds.}{RA}
            \mess{RA}{C. Success}{MA}
            \mess{RA}{D. Report Attempts}{MA}
            \mess{MA}{E. Ack.}{RA}
            \prelevel\prelevel\prelevel\prelevel\prelevel
            
            \mess{TSM}{1. STCP}{TA}
            \mess{TSM}{2. Reveal $C$}{TA}
            \node[below=3mm,anchor=east](t1) at (mess to) {3. Unseal $C$};
            \mess{TA}{4. Show $C$}{TSM}
            \mess{TSM}{5. STCP}{RA}
            \mess{TSM}{6. Lookup $C$}{RA}
            \mess{RA}{7. Revoked $RC$}{TSM}
            \mess{TSM}{8. Ack.}{RA}
            \mess{TSM}{9. Revoke $RC$}{TA}
            \node[below=3mm,anchor=east](t1) at (mess to) {\parbox[b]{1.9cm}{\raggedleft10. Update $RC \in C$}};
            \mess{TA}{11. Success}{TSM}
        \end{sequencediagram}
	}
	\caption{Proposed credential revocation. \emph{Steps A--E are independent of 1--11.}}
	\label{fig:revoc}
	\vspace{-0.5cm}
\end{figure}

Privacy-preserving credential schemes, e.g. DAA and anonymous credentials, are beneficial in verifying credentials without divulging or linking users' identities.   However, to serve as an initial baseline, we scope the focus of this work is scoped to centralised deployments for applications where the concern of violating credential privacy has far fewer consequences than government electronic ID cards or TPMs on consumer devices. As such, we consider Industrial IoT (IIoT), logistics, and public devices in smart cities to serve as three potential application domains.   Relaxing this constraint provides us with headway to pursue simpler, PKI-based solutions as a first step for providing TEE credential revocation with mutual attestation.  We suggest two approaches for blacklisting and whitelisting using a trusted RA, the procedure for which illustrated in Figure~\ref{fig:revoc} and described as follows:
\begin{enumerate}
\item  $TSM$ and $TA$ form a STCP using mutual remote attestation to verify each platforms' integrity and bootstrap a secure channel.
\item[2--3.] $TSM$ instructs $TA$ to reveal the current credentials in use, $C$, which are then unsealed from storage, e.g. encrypted in untrusted storage or an SE.
\item[4.] $C$ is transmitted to the $TSM$ over the STCP by $TA$.
\item[5.] The $TSM$ forms a STCP with the revocation authority, $RA$, who maintains the master revocation list of white- or blacklisted credentials.
\item[6--8.] $TSM$ submits $C$ to RA, who returns a list of the revoked credentials in $C$, i.e. $RC \subseteq C$, from its master revocation list (MRL).
\item[9--11.] $TSM$ instructs $TA$ to revoke $RC$ internally; $TA$ performs the deletion and acknowledges its success.
\end{enumerate}

Note that a malicious device may purposefully fail to update the status of $RC$ internally and attempt to reuse revoked credentials.  Consequently, the use of revoked credentials should be reported to $MA$ responsible for decommissioning compromised devices, a simple protocol for which is listed in Steps A--E in Figure \ref{fig:revoc} based on mutual attestation involving $(MA,RA)$.  Like~\cite{lueks}, delegating revocation list management to RA removes the burden of potentially multiple verifiers synchronising a single list; the TSM can submit a lookup request to the RA, who queries the blacklist or whitelist in $O(1)$ using an associative array.    Either black- or whitelisting can be performed in this model. For blacklisting, the RA maintains a master revocation list (MRL) of revoked credentials that should not be used in any transaction in which the MA submits credentials it wishes to revoke to RA (\emph{maintainer-instantiated revocation}).  Here, RA tests the revocation status of $C$ by verifying $C \notin MRL$.   Whitelisting, conversely, comprises a list of only the permitted credentials; a credential is revoked by removing it from the whitelist and, if applicable, updating the list with its replacement.  Revocation is tested by verifying $C \in MRL$. 

\section{Remote Backups}
\label{sec:backup}


Backup is the process of securely retrieving the set of credentials, $C$, belonging to a TA for remote storage.  In standard practice, backups underpin disaster recovery plans  -- as stipulated by ISO 27001:2013~\cite{iso} -- for recovering data from corruption and accidental deletion.  Backups may also constitute part of a data retention policy, where device data is used as evidence of regulatory adherence.  Secure backup is beneficial when the original credential has non-trivial reinitialisation costs. Next, we examine related work in the backup of remote credentials aboard secure and trusted execution technologies.

Kostiainen et al.~\cite{kost1} address TEE credential backup, restoration and disabling on consumer mobile phones, proposing two solutions.  The first uses a SE -- a SIM card -- in which the TEE credentials are protected under a SIM-specific key provisioned by its provider.  This allows the user to uninstall a familiar hardware element, i.e. the SIM, before releasing the device for repairs or to lend to an untrusted user.  On reinserting the SIM, an on-board TEE credential manager is used to decrypt and re-initialise the encrypted credentials.  The second solution involves the use of a removable microcontroller to counter an honest-but-curious remote server, $RS$.  $RS$ possesses a shared key $K_s$ with the TEE, and stores the backups using a secure counter for rollback protection.  To prevent $RS$ reading the credentials, the TEE encrypts them under a separate key, $K$, derived from a local counter on the microcontroller, and re-encrypts them under $K_s$.

Akram et al.~\cite{akram:lost} examine credential restoration for multi-application smart cards on smartphone SEs.  The SE's Trusted Environment and Execution Manager (TEM), which dynamically enforces the smart card's run-time security policies, is expanded to facilitate credential backups and restoration.  A Backup and Restoration Manager (BRM) is added to the smart card software stack that interfaces with a TEM-resident backup token handler, which stores tokens issued by application service providers.  The user first registers the BRM with a backup server (BS) and, when the user wishes to perform the backup, the BRM encrypts the token(s) and communicates them to BS.   To restore data, such as to a new card, the user provides the BRM with his/her BS credentials to download the backed-up tokens.  A secure channel is formed and the token(s) authenticate the credential restoration from the service provider(s).

\subsection{Proposed High-level Backup Procedure}
\label{sec:backupprotocol}

\begin{figure}
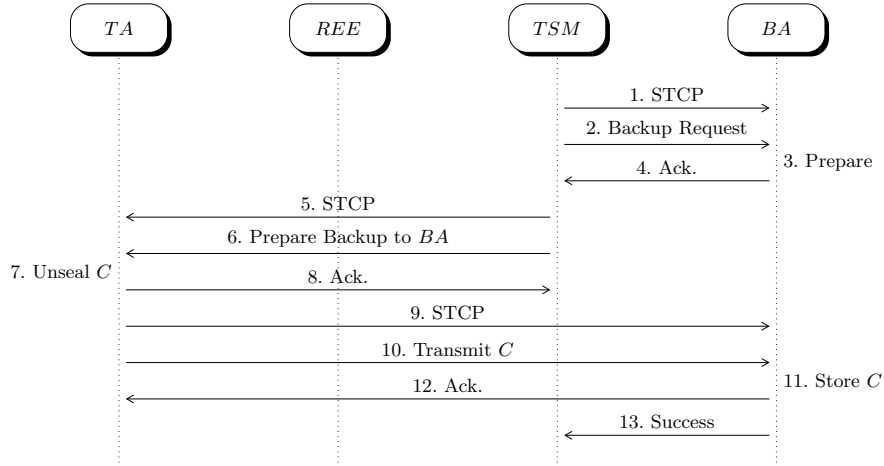

	\centering
	\resizebox{\linewidth}{!}{%
		\begin{sequencediagram}
		    \tikzstyle{inststyle}+=[rounded corners=3mm]
		    \newinst{TA}{$TA$}
            \newinst[2]{REE}{$REE$}
			\newinst[2]{TSM}{$TSM$}
			\newinst[2]{BA}{$BA$}
            \mess{TSM}{1. STCP}{BA}
            \mess{TSM}{2. Backup Request}{BA}
            \node[below=3mm,anchor=west](t1) at (mess to) {3. Prepare};
            \mess{BA}{4. Ack.}{TSM}
            \mess{TSM}{5. STCP}{TA}
            \mess{TSM}{6. Prepare Backup to $BA$}{TA}
            \node[below=3mm,anchor=east](t1) at (mess to) {7. Unseal $C$};
            \mess{TA}{8. Ack.}{TSM}
            \mess{TA}{9. STCP}{BA}
            \mess{TA}{10. Transmit $C$}{BA}
            \node[below=3mm,anchor=west](t1) at (mess to) {11. Store $C$};
            \mess{BA}{12. Ack.}{TA}
            \mess{BA}{13. Success}{TSM}
        \end{sequencediagram}
	}
	\caption{Proposed high-level remote credential backup protocol.}
	\label{fig:backup}
\end{figure}

We introduce a trusted Backup Authority (BA) responsible for storing retrieved credentials.  This may be a cloud-based storage provider or Hardware Security Module (HSM) possessed by the credential issuing authority; the precise means by which BA securely stores credentials is out-of-scope in this work. Importantly, we note that credential restoration can be performed by executing the proposed Remote Update procedure in Section~\ref{sec:update}. The proposed procedure between the target TA and BA is shown in Figure~\ref{fig:backup}, which is described as follows:
\begin{enumerate}
\item[1-4.] $TSM$ and $BA$ establish a secure and trusted channel with mutual attestation, and $TSM$ requests $BA$ to prepare for backup.
\item[5-6.] $TSM$ forms an STCP with $TA$ and commands it prepare the credentials for remote backup; $TSM$ also provides the identity of $BA$.
\item[7-8.] $TA$ unseals the credentials to transmit to $BA$, and notifies $TSM$ that they were unsealed successfully.
\item[9-12.] $TA$ and $BA$ form a STCP over which $C$ is transmitted to and stored securely by $BA$.  While $TSM$ is considered trusted, a direct connection between $TA$ and $BA$ mitigates the risk of unnecessary credential exposure to $TSM$.
\item[13.] $BA$ notifies $TSM$ that $C$ from $TA$ was backed-up successfully.
\end{enumerate}

\section{Remote Updates}
\label{sec:update}

Remotely updating credentials is beneficial during routine renewal schedules; for example, with X.509 certificates that reach their validity expiry date, or the device is relocated and the organisational unit to which the credential is issued is no longer valid.  Generally speaking, update is the process by which an outdated credential, $c_i$, is securely replaced by a freshly issued $c'_{i}$.  Once replaced, $c_i$ should be revoked to prevent reusing obsolete credentials.   Little work has been conducted regarding TEE credential updates, which is likely due to the simplicity of a TSM transmitting a new credential over a secure channel or, indeed, the similarity with backup restoration (Section \ref{sec:backup}).  Updates can be considered a variation of backup restoration where $c'_i$ is retrieved from an update server; the addition is revoking $c_i$, achievable using the revocation process proposed in Section~\ref{sec:revoc}.  Next, we describe how this can be achieved with mutual attestation.

\subsection{Proposed High-level Update Procedure}


We reintroduce the maintenance authority (MA) from Section~\ref{sec:revoc}, which issues credential updates as part of a standard rotation policy.  If desired, the MA is also responsible for registering obsolete credentials with the revocation authority.  The high-level update mechanism is as follows:

\begin{figure}
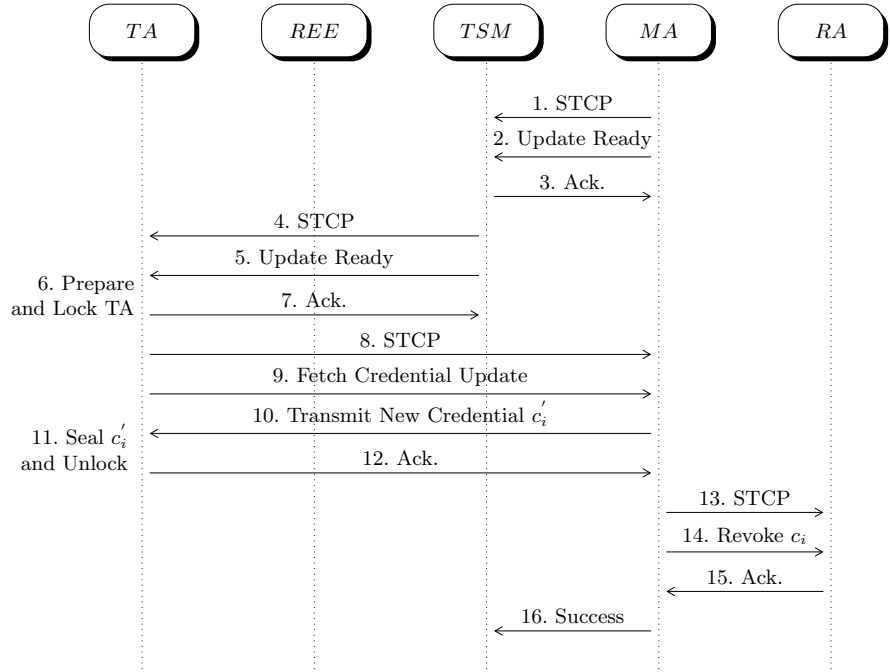

	\centering
	\resizebox{\linewidth}{!}{%
		\begin{sequencediagram}
		    \tikzstyle{inststyle}+=[rounded corners=3mm]
		    \newinst{TA}{$TA$}
            \newinst[1]{REE}{$REE$}
			\newinst[1]{TSM}{$TSM$}
			\newinst[1]{MA}{$MA$}
			\newinst[1]{RA}{$RA$}
            \mess{MA}{1. STCP}{TSM}
            \mess{MA}{2. Update Ready}{TSM}
            \mess{TSM}{3. Ack.}{MA}
            \mess{TSM}{4. STCP}{TA}
            \mess{TSM}{5. Update Ready}{TA}
            \node[below=3mm,anchor=east](t1) at (mess to) {\parbox[b]{1.9cm}{\raggedleft6. Prepare and Lock TA}};
            \mess{TA}{7. Ack.}{TSM}
            \mess{TA}{8. STCP}{MA}
            \mess{TA}{9. Fetch Credential Update}{MA}
            \mess{MA}{10. Transmit New Credential $c^{'}_i$}{TA}
            \node[below=2mm,left=1mm,anchor=east](t1) at (mess to) {\parbox[b]{1.7cm}{\raggedleft11. Seal $c^{'}_i$ and Unlock}};
            \mess{TA}{12. Ack.}{MA}
            \mess{MA}{13. STCP}{RA}
            \mess{MA}{14. Revoke $c_i$}{RA}
            \mess{RA}{15. Ack.}{MA}
            \mess{MA}{16. Success}{TSM}
        \end{sequencediagram}
	}
	\caption{Proposed credential update procedure.}
	\label{fig:update}
	\vspace{-0.5cm}
\end{figure}

\begin{enumerate}
\item $MA$, who provides the updated credentials, establishes an STCP with $TSM$.
\item[2--3.] $MA$ notifies the $TSM$ of an updated credential.  This may include identities of which TEEs need updated or all TEEs.
\item[4--5.] An STCP is conducted between $(TSM, TA)$ and an update preparation command is transmitted to $TA$, along with an optional ID of MA from whom to retrieve the update.
\item[6.] $TA$ is locked, i.e. prevented from interacting with the REE, until the update is performed in order to prevent the use of outdated credentials.
\item[7.] $TA$ acknowledges to $TSM$ that it is ready to update.
\item[8--10.] $TA$ establishes a STCP with $MA$ to receive the credential update; $MA$ transmits the updated credential, $c'_{i}$, to $TA$.
\item[11--12.] $TA$ seals $c'_i$ to its secure storage for future use; $c_i$ should be deleted internally before unlocking. $TA$ acknowledges that $c'_i$ was initialised successfully.
\item[13--16.] $(MA,RA)$ use an STCP to white- or blacklist the obsolete credential, $c_i$. Lastly, $MA$ acknowledges completion of the update procedure to $TSM$.
\end{enumerate}


\section{Proposed Protocol Analysis}
\label{sec:protocols}

We now formalise the protocols from the high-level procedures presented previously, which are listed in Protocols 1 to 6 using the notation from Table~\ref{tab:notation}.   In Section \ref{sec:reqs}, we outlined the requirements and assumptions of the protocols, which are referred to throughout. 
Each proposed protocol is underpinned by an enhancement of the BTP mutual attestation protocol in \cite{shepherd:ares} for establishing the STCP between the TEEs.   This protocol (Protocol~\ref{protocol:btp}), which establishes a TEE-to-TEE secure channel after exchanging and verifying attestation quotes, is simplified to support authenticated encryption (AE), e.g. AES in GCM mode, rather than a non-AE symmetric scheme with an additional HMAC as in original proposal. This simplification is aimed at reducing protocol implementation complexity and improving performance based on existing benchmarks~\cite{gueron}.   The protocol is based on ephemeral Diffie-Hellman key agreement that achieves session forward secrecy (S2), mutual key establishment (S1) and key freshness (S7).   Moreover, TEE quotes are mutually exchanged for verifying the integrity of each platform, thus satisfying S3 and S4 (mutual trust verification).  The signed attestation values, command instructions, e.g. $Prep\_Backup$ and $Revoke\_Success$, and the shared secret provides mutual entity authentication (S5).   

\begin{algorithm}
	\floatname{algorithm}{Protocol}
	\caption{Proposed Migration Procedure with BTP (MPBT)}
	\label{protocol:mbtp}
	\begin{algorithmic}[1]
		\STATE Execute BTP ($TSM$, $TA_1$)
		\STATE $TSM\rightarrow{TA_{1}}$ : $[(Init\_Migrate$ $||$ $X_{TSM})\sigma_{TSM}]_{AE_{K}}$
        \STATE $TA_1\rightarrow{TSM}$ : $[(TA_{1}\_Ack$ $||$ $X_{TA1})\sigma_{TA1}]_{AE_{K}}$
   
   		\STATE Execute BTP ($TSM$, $TA_2$)
        \STATE $TSM\rightarrow{TA_{2}}$ : $[(Prep\_Migrate$ $||$ $X'_{TSM})\sigma_{TSM}]_{AE_{K'}}$ 
        \STATE $TA_2\rightarrow{TSM}$ : $[(TA_{2}\_Ack$ $||$ $X'_{TA2})\sigma_{TA2}]_{AE_{K'}}$
        \STATE $TSM\rightarrow{TA_{1}}$ : $[(ID_{TA2}$ $||$ $X_{TSM})\sigma_{TSM}]_{AE_{K}}$\\
        
        \STATE Execute BTP ($TA_1$, $TA_2$)
        \STATE $TA_1 \rightarrow{TA_{2}}$ : $[(C$ $||$ $X''_{TA1})\sigma_{TA1}]_{AE_{K''}}$ 
        \STATE $TA_2\rightarrow{TA_1}$ : $[(TA_{2}\_Ack$ $||$ $X''_{TA2})\sigma_{TA2}]_{AE_{K''}}$
        \STATE $TA_2\rightarrow{TSM}$ : $[(TA_{2}\_Done$ $||$ $X'_{TA2})\sigma_{TA2}]_{AE_{K'}}$
        
        \STATE $TSM\rightarrow{TA_{1}}$ : $[(Delete\_Creds$ $||$ $X_{TSM})\sigma_{TSM}]_{AE_{K}}$
        \STATE $TA_1\rightarrow{TSM}$ : $[(TA_{1}\_Success$ $||$ $X_{TA1})\sigma_{TA1}]_{AE_{K}}$
	\end{algorithmic}
\end{algorithm}

\begin{algorithm}
	\floatname{algorithm}{Protocol}
	\caption{Proposed Revocation Lookup with BTP (RLBT)}
	\label{protocol:rltp}
	\begin{algorithmic}[1]
		\STATE Execute BTP ($TSM$, $TA$)
        \STATE $TSM\rightarrow{TA}$ : $[(Reveal\_Creds$ $||$ $X_{TSM})\sigma_{TSM}]_{AE_{K}}$
		\STATE $TA\rightarrow{TSM}$ : $[(C$ $||$ $X_{TA})\sigma_{TA}]_{AE_{K}}$
   
   		\STATE Execute BTP ($TSM$, $RA$)
        \STATE $TSM\rightarrow{RA}$ : $[(Lookup$ $||$ $C$ $||$ $X'_{TSM})\sigma_{TSM}]_{AE_{K'}}$
        \STATE $RA\rightarrow{TSM}$ : $[(Revoked$ $||$ $RC$ $||$ $X'_{RA})\sigma_{RA}]_{AE_{K'}}$
        \STATE $TSM\rightarrow{RA}$ : $[(TSM\_Ack$ $||$ $X'_{TSM})\sigma_{TSM}]_{AE_{K'}}$
        \\ If $RC \neq \emptyset$:
        \STATE $TSM\rightarrow{TA}$ : $[(Revoke$ $||$ $RC$ $||$ $X_{TSM})\sigma_{TSM}]_{AE_{K}}$ 
        \STATE $TA\rightarrow{TSM}$ : $[(Revoke\_Success$ $||$ $X_{TA})\sigma_{TA}]_{AE_{K}}$
	\end{algorithmic}
\end{algorithm}

\begin{algorithm}
	\floatname{algorithm}{Protocol}
	\caption{Proposed Revocation Procedure with BTP (RPBT)}
	\label{protocol:rbtp}
	\begin{algorithmic}[1]
		\STATE Execute BTP ($MA$, $RA$)
        \STATE $MA\rightarrow{RA}$ : $[(Revoke$ $||$ $C$ $||$ $X_{MA})\sigma_{MA}]_{AE_{K}}$
		\STATE $RA\rightarrow{MA}$ : $[(Revoke\_Success$ $||$ $X_{RA})\sigma_{RA}]_{AE_{K}}$
        \STATE If reported credentials ($RepC \neq \emptyset$):
        \\ $RA\rightarrow{MA}$ : $[(Report$ $||$ $RepC$ $||$ $X_{RA})\sigma_{RA}]_{AE_{K}}$
        \\ $MA\rightarrow{RA}$ : $[(Report\_Success$ $||$ $X_{MA})\sigma_{MA}]_{AE_{K}}$
	\end{algorithmic}
\end{algorithm}

\begin{algorithm}
	\floatname{algorithm}{Protocol}
	\caption{Proposed Backup Procedure with BTP (BPBT)}
	\label{protocol:bptp}
	\begin{algorithmic}[1]
		\STATE Execute BTP ($TSM$, $BA$)
		\STATE $TSM\rightarrow{BA}$ : $[(Prep\_Backup$ $||$ $X_{TSM})\sigma_{TSM}]_{AE_{K}}$
        \STATE $BA\rightarrow{TSM}$ : $[(BA\_Ack$ $||$ $X_{BA})\sigma_{BA}]_{AE_{K}}$
   
   		\STATE Execute BTP ($TSM$, $TA$)
        \STATE $TSM\rightarrow{TA}$ : $[(Prep\_Backup$ $||$ $ID_{BA}$ $||$ $X'_{TSM})\sigma_{TSM}]_{AE_{K'}}$
        \STATE $TA\rightarrow{TSM}$ : $[(TA\_Ack$ $||$ $X'_{TA})\sigma_{TA}]_{AE_{K'}}$
        
        \STATE Execute BTP ($TA$, $BA$)
        \STATE $TA\rightarrow{BA}$ : $[(C$ $||$ $X''_{TA})\sigma_{TA}]_{AE_{K''}}$ 
        \STATE $BA\rightarrow{TA}$ : $[(Backup\_Ack$ $||$ $X''_{BA})\sigma_{BA}]_{AE_{K''}}$
        \STATE $BA\rightarrow{TSM}$ : $[(Backup\_Success$ $||$ $X_{BA})\sigma_{BA}]_{AE_{K}}$
	\end{algorithmic}
\end{algorithm}

\begin{algorithm}
	\floatname{algorithm}{Protocol}
	\caption{Proposed Update Procedure with BTP (UPBT)}
	\label{protocol:ubtp}
	\begin{algorithmic}[1]
		\STATE Execute BTP ($MA$, $TSM$)
		\STATE $MA\rightarrow{TSM}$ : $[(Update\_Ready$ $||$ $X_{MA})\sigma_{MA}]_{AE_{K}}$
        \STATE $TSM\rightarrow{MA}$ : $[(TSM\_Ack$ $||$ $X_{TSM})\sigma_{TSM}]_{AE_{K}}$
   
   		\STATE Execute BTP ($TSM$, $TA$)
        \STATE $TSM\rightarrow{TA}$ : $[(Prep\_Update$ $||$ $X'_{TSM})\sigma_{TSM}]_{AE_{K'}}$
        \STATE $TA\rightarrow{TSM}$ : $[(TA\_Ack$ $||$ $X'_{TA})\sigma_{TA}]_{AE_{K'}}$
        
        \STATE Execute BTP ($TA$, $MA$)
        \STATE $TA\rightarrow{MA}$ : $[(Fetch\_Update$ $||$ $X''_{TA})\sigma_{TA}]_{AE_{K''}}$
        \STATE $MA\rightarrow{TA}$ : $[(c'_{i}$ $||$ $X''_{MA})\sigma_{MA}]_{AE_{K''}}$
        \STATE $TA\rightarrow{MA}$ : $[(New\_Cred\_Ack$ $||$ $X''_{TA})\sigma_{TA}]_{AE_{K''}}$
        
        \STATE Execute BTP ($MA$, $RA$)
        \STATE $MA\rightarrow{RA}$ : $[(Revoke$ $||$ $c_{i}$ $||$ $X'''_{MA})\sigma_{MA}]_{AE_{K'''}}$
        \STATE $RA\rightarrow{MA}$ : $[(Revoke\_Success$ $||$ $X'''_{RA})\sigma_{RA}]_{AE_{K'''}}$
        \STATE $MA\rightarrow{TSM}$ : $[(Update\_Success$ $||$ $X_{MA})\sigma_{MA}]_{AE_{K}}$
	\end{algorithmic}
\end{algorithm}

\begin{algorithm}
	\floatname{algorithm}{Protocol}
	\caption{Adapted Bi-directional Trust Protocol (BTP) from~\cite{shepherd:ares}}
	\label{protocol:btp}
	\begin{algorithmic}[1]
		\STATE $A\rightarrow{B}$ : $ID_{A}$ $||$ $ID_{B}$ $||$ $n_{A}$ $||$ $g^{A}$ $||$ $AR_{B}$ \\
        
		\STATE $B\rightarrow{A}$ : $ID_{B}$ $||$ $ID_{A}$ $||$ $n_{B}$ $||$ $g^{B}$ $||$ $\big[(X_{B})\sigma_{B}$ $||$  $(V_{B})\sigma_{B}\big]_{AE_{K}}$ $||$ $AR_{A}$ \\
        $X_{B}$ = $H(ID_{A}$ $||$ $ID_{B}$ $||$ $g^{A}$ $||$ $g^{B}$ $||$ $n_{A}$ $||$ $n_{B})$ \\
        $V_{B}$ = $Q_{B}$ $||$ $n_{B}$ $||$ $n_{A}$
		\STATE $A\rightarrow{B}$ : $[$ $(X_{A})\sigma_{A}$ $||$ $(V_{A})\sigma_{A}$ $]_{AE_{K}}$ \\
        $X_{A}$ = $H(ID_{A}$ $||$ $ID_{B}$ $||$ $g^{A}$ $||$ $g^{B}$ $||$ $n_{A}$ $||$ $n_{B})$ \\
        $V_{A}$ = $Q_{A}$ $||$ $n_{A}$ $||$ $n_{B}$
	\end{algorithmic}
\end{algorithm}

\begin{table}
\caption{Protocol notation.}
\centering
\resizebox{0.66\linewidth}{!}{%
\begin{tabular}{ll} 
\toprule
\textbf{Notation} & \textbf{Description} \\\midrule
$TSM$ & TEE trusted service manager.\\
$BA$ & Credential backup authority.\\
$MA$ & Device maintenance authority.\\
$RA$ & Revocation authority.\\
$TA_{X}$ & TEE trusted application on device $X$.\\
$n_{X}$ & Secure random nonce generated by $X$.\\
$H(D)$ & Secure one-way hash function, $H$, on $D$.\\
$X\rightarrow{Y}$ & Message transmission from $X$ to $Y$.\\
$ID_{X}$ & Identity of X.\\
$A$ $||$ $B$ & Concatenation of $A$ and $B$.\\
$g^{X}$ & Diffie-Hellman exponentiation of $X$.\\
$AR_{X}$ & Attestation request on target entity $X$. \\
$Q_{X}$ & Attestation quote from TEE $X$.\\
$(A)\sigma_{X}$ & \makecell[l]{Signed message $A$ from $X$ under a private-\\public key-pair $(K,P)$.}\\
$[m]_{AE_{K}}$ & \makecell[l]{Message $m$ is encrypted using authenticated\\encryption under session key $K$ derived from\\the protocol's shared secret.} \\
$D'$ & \makecell[l]{Data specific to a separate session to $D$.}\\\bottomrule
\end{tabular}
}
\label{tab:notation}
\end{table}

Crucially, the protocols avoid the use of additional trusted hardware, such as TPMs, secure elements and smart cards (F1).  The protocols are designed to incorporate abstract TAs, which are verified using the quoting abstraction, whether they be Intel SGX enclaves of GP TEE TAs, thus providing TEE agnosticism (F2).  Note, however, that this abstracts away the precision of related work, e.g. Arfaoui et al.~\cite{arfaoui}, which addresses migration specifically in the context of the GP TEE.  Such work incorporates GP TEE-specific entites, such as security domains (SDs) and root SDs, which do not exist on Intel SGX or earlier TPM-based TEEs like Intel TXT~\cite{greene2012intel}.  As such, users of this work should be aware of the implementation specifics when deploying these protocols; we refer users to \cite{arfaoui} and \cite{gp:tmf} for guidance for GP TEEs, and \cite{sgx:costan} for Intel SGX. 

\textbf{Formal Symbolic Verification.}
Scyther by Cremers~\cite{cremers:scyther} was employed to verify the correctness of the proposed protocols.  A protocol is first specified in the Scyther description language, comprising communicating parties (roles), messages and the desired security properties (claims). Scyther verifies whether the protocol specification satisfies those claims under the `perfect cryptography' assumption, whereby an adversary learns nothing from an encrypted message unless the decryption key is known, against all possible behaviours of a Dolev-Yao adversary.  Despite the challenge of security protocol verification being undecidable in general, many practical protocols can be proven correct; notably, Scyther has been used to verify IKEv1 and IKEv2, and the ISO/IEC 9798 authentication protocol family~\cite{cremers:ike}.     
We analyse all protocols using Scyther, testing for the secrecy of transmitted quotes from both parties, e.g. (\texttt{Secret, qta1}) and credentials (\texttt{Secret, c}); aliveness (\texttt{Alive}); replay protection, i.e. non-injective agreement (\texttt{Niagree}) and non-injective synchronisation, (\texttt{Nisynch}), 
 defined in \cite{cremers:scyther}; session key secrecy (\texttt{SKR, K}); and the reachability of all entities, e.g. (\texttt{Reachable, TA}).   We publicly release the protocol specifications for future research by the community (see Section~\ref{sec:intro}).   Scyther found no attacks on any protocol.

\section{Conclusion}
TEEs are emerging as a flexible mechanism for providing a range of assurances regarding the on-device protection of security-critical applications, credentials and related data.  In this work, we presented a suite of proposals for remote TEE credential management using mutual attestation for secure \emph{migration}, \emph{revocation}, \emph{backups}, and \emph{updates}.  After summarising the features of leading TEE implementations, we formalised the threat model, requirements and assumptions for a typical TEE credential deployment in a centralised setting.  Next, we reviewed the state-of-the-art for each credential management challenge, before proposing procedures and protocols for securely realising these notions.  The protocols were formalised and subjected to symbolic verification using Scyther, which found no attacks under the Dolev-Yao adversarial model; the protocol specifications are also published publicly for further research.  In future work, we aim to incorporate privacy-preserving attestation into our protocol suite, which we considered out-of-scope in this work for centralised deployments, through the use of techniques like DAA and Blacklistable Anonymous Credentials (BLACs). We also wish to address decentralised deployments, where devices have intermittent or potentially no access to a centralised TSM.

\subsubsection{Acknowledgements.}
Carlton Shepherd is supported by the EPSRC and the British government as part of the Centre for Doctoral Training in Cyber Security at Royal Holloway, University of London (EP/K035584/1).

%
%
%
%
%

\bibliographystyle{splncs04}
\bibliography{main}

\begin{thebibliography}{10}
\providecommand{\url}[1]{\texttt{#1}}
\providecommand{\urlprefix}{URL }
\providecommand{\doi}[1]{https://doi.org/#1}

\bibitem{akram:lost}
Akram, R.N., Markantonakis, K., Mayes, K.: Recovering from a lost digital
  wallet. In: Embedded and Ubiquitous Computing. pp. 1615--1621. IEEE (2013)

\bibitem{akram:trusted}
Akram, R.N., Markantonakis, K., Mayes, K., Bonnefoi, P.F., Sauveron, D.,
  Chaumette, S.: An efficient, secure and trusted channel protocol for avionics
  wireless networks. In: 35th Digital Avionics Systems Conference. IEEE (2016)

\bibitem{arfaoui}
Arfaoui, G., Gharout, S., Lalande, J.F., Traor{\'e}, J.: Practical and
  privacy-preserving {TEE} migration. In: 9th IFIP International Conference on
  Information Security Theory and Practice. pp. 153--168. Springer (2015)

\bibitem{brickell:epid}
Brickell, E., Li, J.: Enhanced privacy {ID} from bilinear pairing for hardware
  authentication and attestation. International Journal of Information Privacy,
  Security and Integrity  \textbf{1}(1),  3--33 (2011)

\bibitem{chen}
Chen, L., Li, J.: Revocation of {Direct Anonymous Attestation}. In:
  International Conference on Trusted Systems. pp. 128--147. Springer (2011)

\bibitem{sgx:costan}
Costan, V., Devadas, S.: {Intel SGX Explained}. IACR Cryptology ePrint  (2016),
  \url{https://eprint.iacr.org/2016/086.pdf}

\bibitem{cremers:scyther}
Cremers, C.: The {Scyther} tool: Verification, falsification, and analysis of
  security protocols. In: Computer Aided Verification. pp. 414--418. Springer
  (2008)

\bibitem{cremers:ike}
Cremers, C.: Key exchange in {IPsec} revisited: Formal analysis of {IKEv1} and
  {IKEv2}. In: European Symposium on Research in Computer Security. Springer
  (2011)

\bibitem{gp:tee}
{GlobalPlatform}: {TEE Protection Profile (v1.2)} (2014)

\bibitem{gp:tmf}
{GlobalPlatform}: {TEE Management Framework (TMF), v1.0} (2016)

\bibitem{greene2012intel}
Greene, J.: Intel {Trusted eXecution Technology (TXT)}: Hardware-based
  technology for enhancing server platform security. Tech. rep., Intel, Inc.
  (2012)

\bibitem{greveler}
Greveler, U., Justus, B., Loehr, D.: Mutual remote attestation: Enabling system
  cloning for {TPM}-based platforms. In: International Workshop on Security and
  Trust Management. pp. 193--206. Springer (2011)

\bibitem{gueron}
Gueron, S.: {AES-GCM} for efficient authenticated encryption -- ending the
  reign of {HMAC-SHA-1}. In: Real World Cryptography (2013)

\bibitem{iso}
{International Standards Organisation}: {ISO 27001:2013 -- Information Security
  Management} (2013), \url{https://www.iso.org/standard/54534.html}

\bibitem{katz}
Katzenbeisser, S., Kursawe, K., Stumpf, F.: Revocation of {TPM} keys. In:
  International Conference on Trusted Computing. pp. 120--132. Springer (2009)

\bibitem{kost}
Kostiainen, K., Asokan, N., Afanasyeva, A.: Towards user-friendly credential
  transfer on open credential platforms. In: Applied Cryptography and Network
  Security. Springer (2011)

\bibitem{kost1}
Kostiainen, K., Asokan, N., Ekberg, J.E.: Credential disabling from trusted
  execution environments. In: 15th Nordic Conference on Secure IT Systems.
  Springer (2010)

\bibitem{lueks}
Lueks, W., Alp{\'a}r, G., Hoepman, J.H., Vullers, P.: Fast revocation of
  attribute-based credentials for both users and verifiers. Computers \&
  Security  \textbf{67} (2017)

\bibitem{sadeghi2015security}
Sadeghi, A.R., Wachsmann, C., Waidner, M.: Security and privacy challenges in
  industrial internet of things. In: 52nd Design Automation Conference. ACM
  (2015)

\bibitem{shepherd:ares}
Shepherd, C., Akram, R.N., Markantonakis, K.: Establishing mutually trusted
  channels for remote sensing devices with trusted execution environments. In:
  12th International Conference on Availability, Reliability and Security. ACM
  (2017)

\bibitem{shepherd2017towards}
Shepherd, C., Akram, R.N., Markantonakis, K.: Towards trusted execution of
  multi-modal continuous authentication schemes. In: Proceedings of the 32nd
  Symposium on Applied Computing. pp. 1444--1451. ACM (2017)

\bibitem{shepherd:secure}
Shepherd, C., Arfaoui, G., Gurulian, I., Lee, R.P., Markantonakis, K., Akram,
  R.N., Sauveron, D., Conchon, E.: Secure and trusted execution: Past, present,
  and future -- a critical review in the context of the {Internet of Things}
  and {Cyber-Physical Systems}. In: 15th IEEE International Conference on
  Trust, Security and Privacy in Computing and Communications. pp. 168--177.
  IEEE (2016)

\end{thebibliography}
\end{document}